# Absorption line shape recovery beyond the detection bandwidth limit: application to the precision spectroscopic measurement of the Boltzmann constant

F. Rohart[1], S. Mejri[2,3], P.L.T. Sow[2,3], S.K. Tokunaga[3,2], C. Chardonnet[2,3], B. Darquié[2,3], H. Dinesan[4], E. Fasci[4], A. Castrillo[4], L. Gianfrani[4] and C. Daussy[3,2]*

[1] *Laboratoire de Physique des Lasers, Atomes et Molécules, Université de Lille 1, F-59655 Villeneuve d'Ascq cedex, France*
[2] *CNRS, UMR 7538, Laboratoire de Physique des Lasers, F-93430 Villetaneuse, France*
[3] *Université Paris 13, Sorbonne Paris Cité, Laboratoire de Physique des Lasers, F-93430 Villetaneuse, France*
[4] *Dipartimento di Matematica e Fisica, Seconda Università di Napoli, Viale Lincoln 5, I-81100 Caserta, Italy*

*Author, to whom any correspondence should be addressed.
E-mail: christophe.daussy@univ-paris13.fr

A theoretical model of the influence of detection bandwidth properties on observed line shapes in laser absorption spectroscopy is described. The model predicts artificial frequency shifts, extra broadenings and line asymmetries which must be taken into account in order to obtain accurate central frequencies and other spectroscopic parameters. This reveals sources of systematic effects most probably underestimated so far potentially affecting spectroscopic measurements. This may impact many fields of research, from atmospheric and interstellar physics to precision spectroscopic measurements devoted to metrological applications, tests of quantum electrodynamics or other fundamental laws of nature. Our theoretical model is validated by linear absorption experiments performed on $H_2O$ and $NH_3$ molecular lines recorded by precision laser spectroscopy in two distinct spectral regions, near- and mid-infrared. Possible means of recovering original line shape parameters or experimental conditions under which the detection bandwidth has a negligible impact, given a targeted accuracy, are proposed. Particular emphasis is put on the detection bandwidth adjustments required to use such high-quality molecular spectra for a spectroscopic determination of the Boltzmann constant at the 1 ppm level of accuracy.

## I. INTRODUCTION

Spectroscopy, the study of the interaction between radiated energy and matter, has played and is playing a crucial role in many domains of Science. One can think of the development of quantum physics and the study of atomic structure. In physical and analytical chemistry, atomic and molecular spectra are used to detect, identify and quantify key information. In astronomy spectra observed with telescopes are used to determine the chemical composition and physical properties of astronomical objects.

In recent years, there has been a growing interest in highly precise and accurate observations of the shape of molecular spectral lines. Climate modeling and global change research programs are setting unprecedented accuracy targets in gas-sensing missions for atmospheric $CO_2$ and other greenhouse gases [1]. The desired uncertainty can only be reached by making high quality measurements of spectroscopic parameters (including pressure broadening coefficients and line intensity factors) using well-designed and characterized experiments [2-5]. Similarly, the interpretation of spectra of astrophysical and planetary interest needs a precise knowledge of line-shape parametres and current databases require further improvements [6, 7]. The accurate observation of line profiles is also crucial to the physics of collisions by testing theories modeling the shape of atomic or molecular transitions perturbed by collisions [8-12]. Highly accurate spectroscopy plays as well a decisive role in precision measurements devoted to metrological applications and tests of fundamental physics. Spectroscopy of constituant of atoms, atoms themselves, molecules, simple exotic atoms (anti-hydrogen, muonic atoms,…) is being used to test quantum electrodynamics, to test fundamental symetries (such as P, PT, CPT, matter/antimatter,…), to test postulates of quantum mechanics (symmetrization postulate, wavefunction collapse,…). It is being used also to measure fundamental constants and properties (fine structure constant, Rydberg constant, proton-to-electron mass ratio, Boltzmann constant,…) and their possible variation in time [13, 14].

The present paper rises a source of systematic effects that may have been underestimated so far: it investigates the possible inaccuracy in experimental determination of line shape parameters – central frequencies, diverse contributions to the width, line intensities,… – due to the unavoidably limited bandwidth of detection chains with which spectra are recorded. This study was initially motivated by ongoing experiments dedicated to Doppler Broadening Thermometry (DBT), a relatively new technique used in our laboratories to determine the Boltzmann constant $k_B$ [15, 16]. DBT consists in retrieving the Doppler width from the accurate measurement of the linear absorption profile of an atomic or a molecular transition, in a gaseous sample at the thermodynamic equilibrium. A determination of $k_B$ by DBT with a combined uncertainty of 1 ppm, comparable to the best current uncertainty obtained using acoustic methods, would make a significant contribution to any new value of this constant determined by the Committee on Data for Science and Technology (CODATA). Furthermore, having multiple independent measurements at these accuracies opens the possibility of defining the kelvin by fixing $k_B$, an exciting prospect considering the upcoming redefinition of the International System of Units (SI) [17- 19]. One important result of this work is to quantify the inaccuracy on the determination of $k_B$ due to the limited detection bandwidth.

This paper begins by describing in Section II a model showing that a finite measurement bandwidth results in a distortion of the line shape accompanied by a shift of the retrieved central frequency. Section III validates this model by the analysis of high-quality linear absorption molecular spectra recorded in different spectral regions. Linear absorption is a particularly interesting and challenging case as it puts at stake complex profiles involving many inhomogeneous and homogeneous contributions to shifts, broadenings and narrowings of the line. Section IV describes possible means of recovering the original central frequency and line shape parameters. Finally section V discusses the implications of these effects for precision measurements of the Doppler width by DBT.

## II. THEORETICAL MODEL

Two key cases are considered. First, we consider the effect of the detection bandwidth on the recorded absorption signal assuming a continuous sweep of the laser frequency. We then extend this treatment to a frequency sweep consisting of a serie of discrete steps, as is the case in most spectrometers.

## A. Impact of the detection bandwidth on recorded absorption signals

Let us consider the case of an isolated line, the absorbance $A(\nu)$ of which is small enough so that the Beer-Lambert law can be considered in its linear form (sample transmittance is given by $1-A(\nu)$)[1]. In laser absorption spectroscopy, the absorbance $A(\nu)$ is typically recorded as a function of the laser frequency $\nu$ that evolves linearly with time $t$ at a constant sweep speed $\dot{\nu} = d\nu/dt$ (in Hz/s). The detection chain can be viewed as a low-pass filter, generally of either the first or the second order with -6 or -12 dB/oct roll-off, respectively.

In case of the 2$^{nd}$ order filter, frequently considered when using a lock-in detection, the time evolution of the recorded signal $D(t)$ is given by [21]

$$\tau_D^2 \frac{d^2 D(t)}{dt^2} + \frac{\tau_D}{Q} \frac{dD(t)}{dt} + D(t) = A(t) \qquad (1)$$

where $\tau_D$ is the filter time constant, $Q$ the filter quality factor which is usually less than 1 for useful low-pass filters and $A(t)$ is the time evolution of the absorbance $A(\nu)$ under study.
Eq. (1) can be re-written in terms of laser frequency as

$$Q^2 \nu_D^2 \frac{d^2 D(\nu)}{d\nu^2} + \nu_D \frac{dD(\nu)}{d\nu} + D(\nu) = A(\nu) \qquad (2)$$

where $\nu_D = \tau_D \dot{\nu}/Q$ is a "frequency constant", a characteristic parameter of experimental recording conditions, and $D(\nu)$ is the signal recorded at the laser frequency $\nu$.
In order to avoid unwanted signal deformations, the time constant $\tau_D$, and therefore $\nu_D$, has to be chosen to be small enough with respect to the time interval required to record the whole line profile [22]. In this case, by analogy with signal processing theory [21, 23], an approximated solution of Eq. (2) is a detected signal $D(\nu)$ that reproduces the expected absorbance signal $A(\nu)$ with a slight frequency lag given approximately by the frequency constant $\nu_D$.

In order to get a better insight into the actual recorded signal, a more accurate solution can be obtained from a development of $D(\nu)$ in powers of $\nu_D$ and written as

$$D(\nu) = A(\nu - \nu_D) + \sum_{k=1}^{\infty} (\nu_D)^k . \alpha_k(\nu) \qquad (3)$$

where $\alpha_k(\nu)$ are trial functions and $A(\nu - \nu_D)$ can be expressed as a function of $A(\nu)$ derivatives

$$A(\nu - \nu_D) = A(\nu) + \sum_{\ell=1}^{\infty} \frac{(-\nu_D)^\ell}{\ell!} . \frac{d^\ell A(\nu)}{d\nu^\ell} . \qquad (4)$$

By using Eqs. (3) and (4) and comparing identical powers of $\nu_D$, Eq. (2) leads to a set of coupled differential equations linking $\alpha_k(\nu)$ functions and their 1$^{st}$ and 2$^{nd}$ derivatives to $A(\nu)$ derivatives. After simple calculations, one obtains:

$$D(\nu) = A(\nu - \nu_D) + (1/2 - Q^2).(\nu_D)^2 . \frac{d^2 A(\nu)}{d\nu^2} - (5/6 - 2Q^2).(\nu_D)^3 . \frac{d^3 A(\nu)}{d\nu^3} + \ldots . \qquad (5)$$

---

[1] In order to avoid any coherent transient problem due to rapid passage, which is beyond the scope of this paper [20], collisional relaxation is assumed large enough so that $A(\nu)$ actually corresponds to the steady state regime at the frequency $\nu$. Modifications of the laser frequency at the scale of one relaxation time constant must be smaller than the collisional line width, i.e. $\dot{\nu} \ll 2\pi \Delta \nu_{coll}^2$.

To lowest order, the recorded line is thus simply shifted (to higher or lower frequencies depending on whether the laser frequency is increasing or decreasing) without any deformation. It is worth noting that this shift, given by the frequency constant $\nu_D$, is independent of the actual line shape. The second term depends on the second derivative of $A(\nu)$. In case of a symmetric line shape, this term reaches a negative minimum value at the line center frequency $\nu_0$ and is nearly zero in the vicinity of the two frequencies corresponding to half maximum. This entails a reduction of the line amplitude and thus a modification of the line shape resulting in an asymmetry and an increased line width. However, it is interesting to note that this term vanishes for $Q = \sqrt{1/2}$, a condition which is fulfilled in the case of the so-called Butterworth filter [21]. Unfortunately, it seems that for many lock-in amplifiers [24], the -12dB/oct roll-off is obtained with two identical successive 1$^{st}$ order filters, where $Q = 1/2$ instead. The third term depends on the third derivative of $A(\nu)$, an odd function when the line shape is symmetric. So this term contributes also to an apparent line shape asymmetry. Subsequent terms of Eq. (5) are generally small and contribute to more complex line shape distortions. Thus in conclusion, in the case of a 2$^{nd}$ order low pass filter, we expect an impact on the line width as well as a distorted line shape which leads to an apparent shift on the line center frequency.

This description can be extended to the case of a 1$^{st}$ order filter. It is easily shown that previous equations also hold in the case of $Q = 0$, except the frequency constant $\nu_D$ is given by $\nu_D = \tau_D . \dot{\nu}$, where $\tau_D$ is the 1$^{st}$ order filter time constant.

**B. Impact of the detection bandwidth on recorded dipole moment correlation function**

By moving to the time domain by Fourier transform, the signal $D(\nu)$ can be replaced by its Fourier transform $\tilde{D}(\tau)$:

$$\tilde{D}(\tau) = \int_{-\infty}^{+\infty} D(\nu) . \exp(+2\pi i \nu \tau) . d\nu \quad . \tag{6}$$

A similar equation holds for the Fourier transform of the absorbance $A(\nu)$ which can be interpreted as proportional to the correlation function $\Phi(\tau)$ of the dipole moment induced in the gaseous medium after a laser excitation starting at time $\tau = 0$ [12, 25, 26]. For the 2$^{nd}$ order filter case, Eq. (2) becomes [21][2]:

$$\tilde{D}(\tau) \propto \tilde{G}(\tau) . \Phi(\tau) \tag{7}$$

where $\tilde{G}(\tau)$ can be seen as the complex linear "gain" of the system given by

$$\tilde{G}(\tau) = \frac{1}{1 - 2\pi i \nu_D \tau - (2\pi Q \nu_D \tau)^2} \quad . \tag{8}$$

Eq. (7) tells us that the detection system properties can be taken into account by simply multiplying the correlation function $\Phi(\tau)$, by the gain $\tilde{G}(\tau)$ of the system, thus leading to the following expression:

$$\tilde{D}(\tau) \propto \frac{\Phi(\tau)}{1 - 2\pi i \nu_D \tau - (2\pi Q \nu_D \tau)^2} \tag{9}$$

that can be included in the spectrometer instrumental function in the line shape modeling.

---

[2] Note that, in contrast to usual signal treatment theory, frequency and time domains are exchanged, $\nu$ and $\tau$ (the latter not to be confused with time $t$) being conjugate variables.

It is worth noting that this result is exact, quite an interesting property for fitting purposes by comparison with the frequency domain solution given by Eq. (5) development. Moreover, $\Phi(\tau)$ has an analytical form for a number of line profiles such as Voigt, Galatry, Speed-Dependent Voigt and Speed-Dependent Galatry models [12, 26-29], including the case of frequency modulated spectral profiles [30]. Of course, the result is not restricted to the 2$^{nd}$ order low pass filters (or to the 1$^{st}$ order ones by setting $Q = 0$), and can be generalized to more complicated detection schemes.

This result can be extended to optically thick media for which the Beer-Lambert law cannot be replaced by its linear approximation. In this latter case, the sample transmittance $T(\nu)$ will be given by $T(\nu) = \exp[-A(\nu)]$ and its Fourier transform $\tilde{\Theta}(\tau)$ [31]. Then, the sample detected transmission will be proportional to the Fourier transform of $\tilde{G}(\tau).\tilde{\Theta}(\tau)$.

### C. Application to the Voigt profile

As an illustrative example, let us consider the Voigt profile in more detail. The corresponding dipole correlation function is [12, 29]

$$\Phi_{Voigt}(\tau) \propto \exp\left[2\pi(i\nu_0 - \Delta\nu_{coll})\tau - (\pi\Delta\nu_{Dop}\tau)^2\right] \qquad (10)$$

where $\nu_0$ is the line center frequency, $\Delta\nu_{coll}$ the collisional width (half width at half maximum, HWHM, of the Lorentzian contribution to the profile) and $\Delta\nu_{Dop}$ the Doppler width (half-width at $1/e$ of the maximum of the Gaussian contribution to the profile). By using the Taylor expansion of $\ln[\tilde{G}(\tau)]$ along with Eq. (10), the Fourier transform $\tilde{D}_{Voigt}(\tau)$ of the detected signal $D(\nu)$ becomes:

$$\tilde{D}_{Voigt}(\tau) \propto \exp\left\{ 2\pi\left[i(\nu_0 + \nu_D) - \Delta\nu_{coll}\right]\tau \right.$$
$$\left. -\left[\Delta\nu_{Dop}^2 + (2-4Q^2)\nu_D^2\right](\pi\tau)^2 - i(1/3 - Q^2).(2\pi\nu_D\tau)^3 + ... \right\}. \qquad (11)$$

For small values of the time $\tau$, this asymptotic expansion done in terms of $\tau$ facilitates useful interpretations, while remaining in good agreement with results from Eq. (5).
The imaginary and real parts of the $\tau$-term show that the absorption line center $\nu_0$ is frequency shifted by $\nu_D$ whereas the collisional broadening $\Delta\nu_{coll}$ remains unaffected.
On the other hand, the $\tau^2$-term is responsible for a modification of the line shape, where the gaussian contribution $\Delta\nu_{Gauss}$ to the Voigt profile appears to be larger than the pure Doppler width, according to the following expression:

$$\Delta\nu_{Gauss} = \sqrt{\Delta\nu_{Dop}^2 + (2-4Q^2)\nu_D^2} \ . \qquad (12)$$

Finally, the $\tau^3$-term is imaginary and thus leads to a line asymmetry. Note that the area of the recorded line remains unaffected since $\tilde{D}(\tau=0) \propto \Phi(\tau=0) = 1$ [21].

At this point, it is useful to introduce some typical values and orders of magnitude to facilitate the discussion about the motivations of our work. Among spectroscopists, a rule of thumb is that the detection time constant $\tau_D$ should be at least 20 times smaller than the duration required for recording the shape over one half-width [22], thereby constraining the detection frequency constant $\nu_D$. For low pressure experiments, namely in the Doppler regime, this condition leads to $\nu_D \approx \Delta\nu_{Dop}/10$ in case of the usual -12 dB/oct roll-off of a 2$^{nd}$ order filter with $Q=1/2$. This means that the recorded line center will be shifted by about a tenth of the Doppler width, a value which could be far from being negligible considering the accuracy and signal to noise ratio achieved with modern laser-based spectrometers.

Under the same operation conditions, the apparent Doppler width will be slightly increased, by about $5\,10^{-3}$. Such a value, actually negligible for common spectroscopic studies, must be considered very carefully in case of the experiments to measure the Boltzmann constant via the Doppler width [32-34], where a 1ppm combined uncertainty is being targeted. In case of a 2$^{nd}$ order filter with the usual quality factor $Q=1/2$ [24], this requirement leads to a $v_D/\Delta v_{Dop}$ ratio as small as 1/700 that requires a very small $\tau_D$ time constant and/or low frequency speed $\dot{v}$. In this respect, it is worth noting the advantage of implementing a 2$^{nd}$ order Butterworth filter in a lock-in amplifier: its specific $Q$-value ($\sqrt{1/2}$) leads to 1$^{st}$ order to a gaussian width which is unaffected by the value of the frequency constant $v_D$ and equal to the Doppler width $\Delta v_{Dop}$ (see Eq. 12).

Note in passing that the theoretical model proposed here can also be applied to line shapes relevant for sub-Doppler (saturated or two-photon absorption) spectroscopy widely used in frequency metrology and high-resolution spectroscopy of molecules, atoms or ions. In these cases, only the homogeneous (Lorentzian) profile needs to be considered.

### D. Step-by-step frequency sweeping mode analysis

So far, we have considered a continuous tuning of the laser frequency. However, for many spectrometers, typical experimental conditions correspond to a frequency $v$ swept step-by-step via a frequency synthesizer, each frequency step $\Delta v$ having a time-duration $\Delta t$, thus leading to an average frequency sweep speed given by $\dot{v} = \Delta v/\Delta t$. As the laser frequency changes, taking the value $v_n$ at time $t_n$, the sample absorbance $A(t)$ should consist of a sequence of steps, each of them being $A(t) = A(v_n)$ for $t_n \leq t < t_n + \Delta t$. After detection, $A(t)$ is treated by the amplifier acting as a low pass filter (Eq. 1), leading to the signal $D(t)$ which is sampled at time $t_n + \Delta t$ (just before a new frequency change), the delay $\Delta t$ allowing for filter integration.

This step-by-step frequency sweeping and sampling mode was modeled by numerically integrating of Eq. (2). This allowed artificial line shapes $D(v)$ to be generated. For simplicity, the gas absorbance $A(v)$ was taken to be a gaussian line shape. In order to measure the distortion caused by the detection system, the 'distorted' $D(v)$ line shapes were fitted using a gaussian profile. The fit parameters, the central frequency $v_{fit}$, the gaussian width $\Delta v_{Gauss}$ and the line amplitude, were then compared to their 'true' values with which the original $A(v)$ profile was generated.

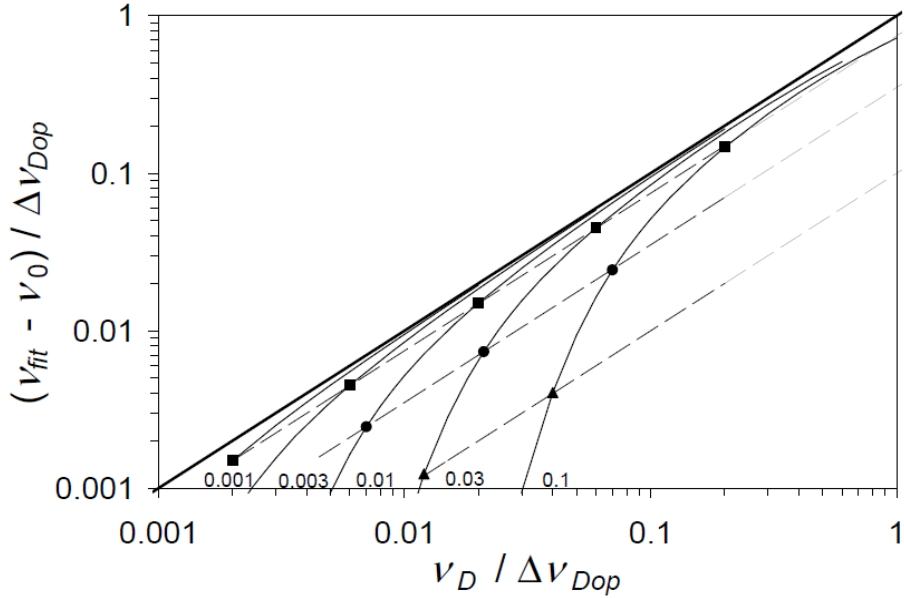

FIG. 1. Influence of frequency sweeping conditions on the line frequency determination (2$^{nd}$ order filter with $Q = 1/2$). Frequency deviations $(\nu_{fit} - \nu_0)/\Delta\nu_{Dop}$ are plotted versus frequency constant $\nu_D/\Delta\nu_{Dop}$ for various frequency steps $\Delta\nu/\Delta\nu_{Dop}$ (ranging from 0.001 up to 0.1 and quoted closed to corresponding full curves). Results corresponding to some $\tau_D/\Delta t$ ratios are drawn in dotted lines and specified by symbols: (▲) 0.2; (●) 0.35; (■) 1.0. The thick line refers to continuous frequency sweeping model.

FIG. 1 shows the deviation of the centre frequency from its original value ($\nu_{fit} - \nu_0$), as a function of the frequency constant $\nu_D$, for a $Q = 1/2$ 2$^{nd}$ order filter. Both axes have been scaled to the Doppler width. Full curves refer to various frequency steps $\Delta\nu$ and dotted lines (specified by symbols) to different values of $\tau_D/\Delta t$. Similar behaviors were obtained for other filters (1$^{st}$ order and 2$^{nd}$ order with $Q = \sqrt{1/2}$) and it is found that the frequency deviations can be written as:

$$\nu_{fit} = \nu_0 + a_\nu \cdot \nu_D, \tag{13}$$

Where $a_\nu$ is a parameter depending on the considered filter (1$^{st}$ order, 2$^{nd}$ order with $Q = \sqrt{1/2}$ or $Q = 1/2$) and on $\tau_D/\Delta t$ ratio. Corresponding $a_\nu$-values are collected in TABLE I.

| $\dfrac{\tau_D}{\Delta t}$ | FILTER | | |
|---|---|---|---|
| | 1$^{st}$ order | 2$^{nd}$ order $Q = 1/2$ | 2$^{nd}$ order $Q = \sqrt{1/2}$ (Butterworth) |
| 0 | 0 | 0 | 0 |
| 0.1 | 0.001 | 0.002 | - |
| 0.2 | 0.036 | 0.098 | - |
| 0.5 | 0.31 | 0.52 | 0.32 |
| 1.0 | 0.58 | 0.75 | 0.65 |
| 2.0 | 0.77 | 0.88 | 0.83 |
| 5.0 | 0.90 | 0.95 | 0.93 |
| ∞ | 1 | 1 | 1 |

TABLE I. $a_\nu$-values (see Eq. 13) computed for various time constant to step duration ratios $\tau_D/\Delta\nu$ and for various detection filters.

Limiting cases are easily understood, $a_\nu$ tending towards zero for small $\tau_D/\Delta t$ values and towards 1 for large $\tau_D/\Delta t$ values. This latter case corresponds to the continuous frequency sweeping operation model which always represents an upper limit for the frequency deviations (see FIG. 1, thick line). The frequency deviation increases with $\nu_D$, whatever the value of $\tau_D/\Delta t$ ratio is, and for a given frequency step value $\Delta \nu$, the smaller is $\tau_D/\Delta t$ the smaller are the frequency deviations. This means that, for each $\Delta t$, a sufficient time interval is left for a perfect signal integration before sampling.

Since for the width, the effects of having a step-by-step spectrometer operation are more complicated, corresponding discussion is postponed to section V.

These models will now be compared to several experiments done at the Laser Physics Laboratory (LPL) of Villettaneuse and at the Molecules and Precision Measurement Laboratory (MPML) of Caserta.

### III. DESIGN FEATURES OF EXPERIMENTAL SETUPS

#### A. Laser Physics Laboratory spectrometer

A set of experiments were done on the saQ(6,3) rovibrational line of the $\nu_2$ vibrational mode of $^{14}NH_3$ recorded at temperature $T \approx 273.15$ K with the LPL's spectrometer operating in the 10 µm range, detailed elsewhere [35, 36]. We simply recall that the frequency was controlled step-by-step and that the signal was amplitude modulated at 40 kHz via a MW-IR frequency mixer and demodulated by a lock-in amplifier (Stanford Research model SR-830) operating in the -12 dB/oct roll-off mode (note that for the current study, line shape broadening due to amplitude modulation is negligible [33]). As explained in Ref. [24], the lock-in amplifier output is a 2$^{nd}$ order system that consists of two successive identical 1$^{st}$ order filters, so our theoretical model is applied with $Q = 1/2$.

Line shapes were recorded at various pressures $P$ (ranging from 0.4 to 4.2 Torr) and using various detection conditions (frequency speeds $\dot{\nu}$ ranging from 5 to 90 MHz/s and indicated amplifier time constants $\tau_D$ from 30 ms to 3 s), so that $\nu_D$ could be varied from 0.3 MHz up to 350 MHz (i.e. from $6\times10^{-3}$ up to 7 Doppler widths) and $\tau_D/\Delta t$ varied from 0.3 to 60.

#### B. Molecules and Precision Measurement Laboratory spectrometer

Experiments were also done with the MPML's 1.39-µm dual-laser spectrometer. The spectrometer has already been described in details elsewhere [34, 37]. For the present study, the intensity of the probe laser was modulated by using a chopper at a frequency of about 2 kHz. Hence, phase-sensitive detection was performed by using a lock-in amplifier (Ametek, model 5209). This amplifier was also operating in the −12 dB/oct roll-off mode. From an analysis of rise times given in Ref. [38], it was concluded that the output filter could be considered as being a 2$^{nd}$ order one with $Q = 1/2$.

Experiments were performed on the $4_{41} \rightarrow 4_{40}$ line of the $H_2^{18}O$ $\nu_1 + \nu_3$ band at the constant temperature of T $\approx$ 296.07 K. Line shapes were recorded at a fixed pressure $P$ (of about 3.6 Torr, from a 97% $^{18}$O-enriched water sample) and using various detection conditions (frequency speeds $\dot{\nu}$ ranging from 10 to 85 MHz/s and indicated amplifier time constants $\tau_D$ from 0.01 to 3 s), so that $\nu_D$ could be varied from 0.5 MHz up to 500 MHz (i.e. from $1.4 \times 10^{-3}$ up to 1.4 Doppler widths) and $\tau_D/\Delta t$ varied from 0.3 to 85.

The large signal to noise ratios achieved allowed an analysis of the influence of spectroscopic parameters on some subtle issues such as deviations from the Voigt profile (see section IV.C)

### C. A representative example: Calibration of the detection system frequency constant

Line shapes of ammonia were recorded with the LPL's spectrometer, under identical thermodynamic conditions, with a $\dot{\nu} = 21.8\,\text{MHz/s}$ sweeping rate and a lock-in amplifier set to a 3 s time constant. This is 1.3 times longer than the time required for sweeping one Doppler half-width ($\approx 49.9$ MHz). Such conditions correspond to a frequency constant $\nu_D = \tau_D.\dot{\nu}/Q = 130$ MHz (about 2.6 Doppler widths). $\tau_D/\Delta t$ was chosen to be large (equal to 44) in order to approach the continuous frequency sweeping mode. The recorded absorption signals are shown in FIG. 2.

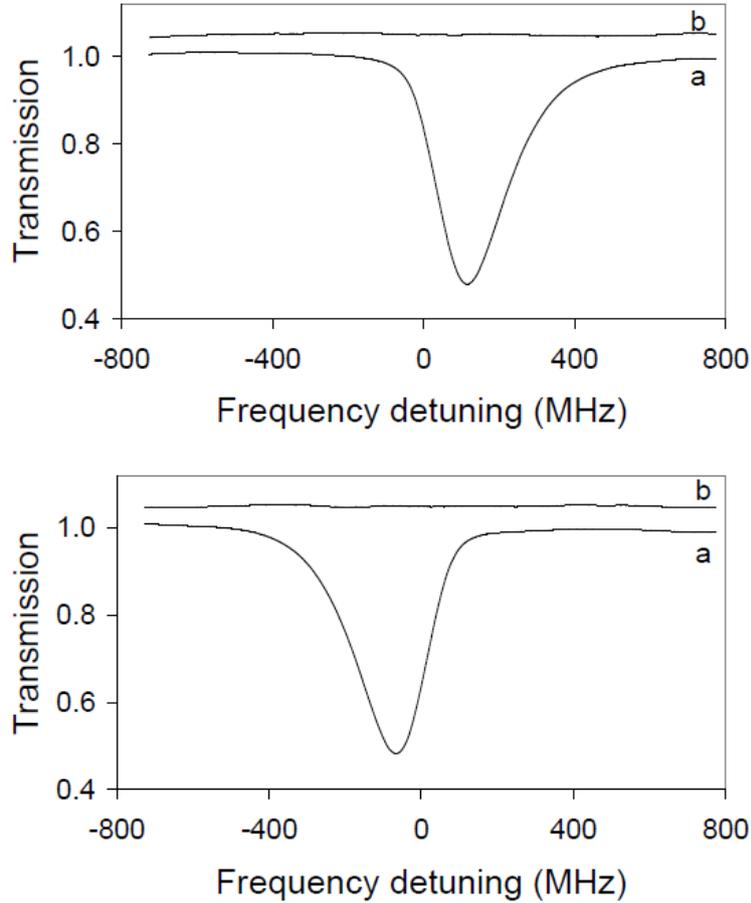

FIG. 2. Transmission of the saQ(6,3) rovibrational line of the $\nu_2$ vibrational mode of $^{14}\text{NH}_3$ recorded for increasing (upper panel) and decreasing (lower panel) frequencies (see text for details). (a): recorded line; (b): unmagnified residuals from a Voigt profile fit taking the detection system properties into account ($\Delta\nu_{col}$ = 13.60(8) and 12.56(7) MHz, respectively, achieved signal to noise ratios $\approx$ 480 and 550, respectively). Frequency scale (in MHz) is shifted by 28.953 600 THz; temperature =273.15 K; pressure $\approx$ 1 Torr; cell length = 4.5 cm; frequency step $\Delta\nu$ = 1.5 MHz; step duration $\Delta t$ = 68.8 ms; indicated lock-in time constant $\tau_D$ = 3 s.

The signals were recorded in both sweep directions, once with increasing and once with decreasing frequencies (shown in the upper and lower panels, respectively). These parameters were chosen specifically to enhance deformations due to the detection system. As a result, the line shapes appear to be strongly asymmetric, exhibiting a sharp leading edge and a soft tail. They also show maximum absorption at frequencies that differ from the true centre by about 90 MHz in each direction, corresponding to a shift of the order of $\nu_D$. However, it is interesting to note the mean value of these two frequencies is in agreement with the frequency obtained by saturated absorption techniques [35].

Both records have been fitted, taking the detection properties into account via an extension of Eq. (9) using a simple Voigt profile including the Beer-Lambert law (see section II.B and C.). In these fits, the central frequency $\nu_{fit}$, the collisional broadening $\Delta\nu_{col}$, the line area and the base line (level and slope) were adjusted, while fixing the Doppler broadening at its theoretical value for T≈273.15 K. Residuals reported in FIG. 2 show that both recorded line shapes are very well reproduced, leading to $\nu_{fit}$ frequencies that differ by 0.33(6) MHz, less than 0.2% of the 180 MHz frequency difference of maximum absorption pikes.

It is important to note that this residual difference can be exactly canceled by setting the lock-in time constant to $\tau_D = 2.993$ s instead of the nominal 3s. The resulting measured central frequency is 28.953 694 15(5) THz, in agreement (to within 2 standard deviations) with the value obtained by saturated absorption techniques, $\nu_0 = 28.953\ 693\ 9(1)$ THz [35]. This shows the importance of an accurate measurement of the lock-in time constant $\tau_D$. This is especially clear from Eqs. (5) or (11) which show that retrieved central frequencies depend linearly on $\tau_D$ via $\nu_D$. Thus, for calibration purposes, we must consider several measurements made at a fixed time constant $\tau_D$ with different ($\Delta\nu_i$, $\Delta t_i$) sweeping conditions. Defining $\nu^i_{fit}$ as the measured central frequency for an experiment $i$, the $\tau_D$ value was adjusted in order to minimize the quantity $\sum_i (\nu^i_{fit} - \nu_0)^2$. This measures the true time constant of the lock-in detection $\tau_D$, which agrees with the nominal values to within 10% or less for large $\tau_D$ and 20% for the smallest ones. These values were then considered for further analysis (sections IV and V). Note that in this investigation, the $Q$-value remains fixed to 1/2, as $Q$- and $\tau_D$-values are strongly correlated via the definition of the frequency constant $\nu_D$.

To conclude, this experiment gives strong support to the above theoretical model and shows that an accurate characterization of the amplifier output filter is required. In particular, the time constant $\tau_D$ must be very well known.

## IV. MOLECULAR RESONANCE FREQUENCY AND LINE SHAPE PARAMETERS RECOVERY

In this section we describe possible means of recovering the original central frequency and line shape parameters of molecular transitions. Spectra have been recorded with LPL's or MPML's spectrometers for which frequency sweeps consist of a serie of discrete steps. Depending on experimental conditions ($\tau_D/\Delta t$ ratio values) either the step-by-step frequency sweeping mode (section II.D) or the continuous frequency sweeping mode approximation has been used (section II.B). As no analytical line shape is presently available for the step-by-step model, line shape fitting procedure has been performed in the frame of the continuous frequency sweeping mode approximation (Eq.(7)), whereas some consequences of the step-by-step frequency sweeping mode have been analysed using numerical simulations (numerical integration of Eq. (2), see section II.D).

## A. Resonance frequency measurement

Several spectra of water vapor were recorded with the MPML's spectrometer in the same thermodynamic conditions, using frequency step $\Delta \nu = 3$ MHz, for increasing values of the frequency constant $\nu_D$. According to section II.D, and considering chosen $\tau_D/\Delta t$ values (from 0.3 to 85), the frequency step influence could not be strictly neglected ($0.25 < a_\nu < 1.0$). In a first approximation, spectra have been fitted by using Eq. (9) in order to test resonance frequency recovery in the frame of the continuous frequency sweeping mode approximation. This was done adopting the suitable profile, a speed dependent Galatry profile, with fixed collisional broadening $\Delta \nu_{col}$, pressure induced frequency line shift, velocity exponent $m$ and the diffusion parameter $\beta_{Gal}$ from ref [34, 37], the Doppler broadening being fixed at its theoretical value (357.05 MHz). In these fits, the central frequency $\nu_{fit}$, the line area and the base line (level and slope) were adjusted. In FIG. 3 the retrieved line-center frequencies were plotted as a function of the frequency constant $\nu_D$.

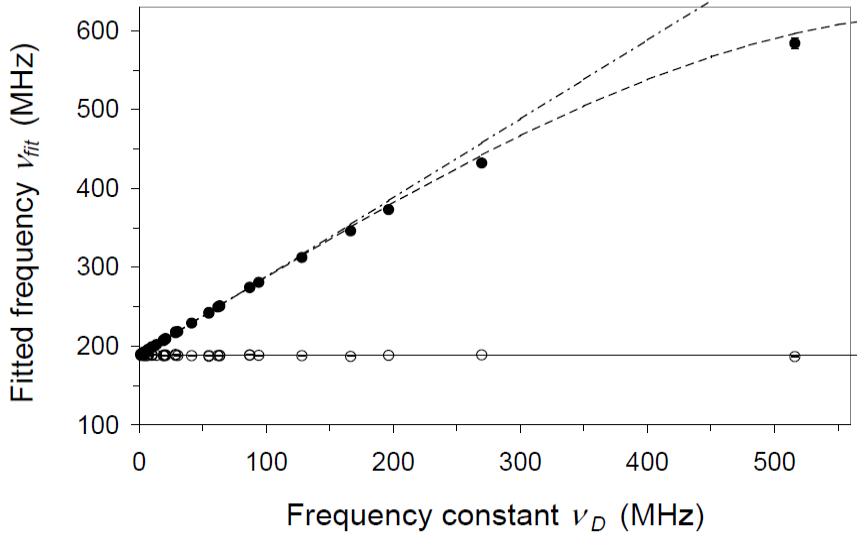

FIG. 3. Absorption frequency of the 7 199.103 cm$^{-1}$ line of H$_2^{18}$O retrieved from a SD-Galatry profile. Fitted central frequencies $\nu$, shifted by 215.823 500 THz, are plotted against the frequency constant $\nu_D$ (see text for details): (•) uncorrected values, (o) values corrected for detection system properties; (mixed line) linear approximation; (dashed line) empirical model; (full line) weighted mean value of corrected data. Error bars are 3 standard deviations.

If detection properties are neglected, one observes a shift of $\nu_{fit}$ proportional to $\nu_D$ for the lower $\nu_D$ values, as expected from Eqs. (5) or (11). For $\nu_D$-values larger than about half the Doppler width, the line asymmetry becomes significant, which explains the observed non linearity. From the retrieved values, this deviation can be well modeled using the empirical expression $\nu_{fit} = \nu_0 + \nu_D \left[1 - 0.125(\nu_D / \Delta \nu_{Dop})^2 \right]$. By contrast, when considering detection properties via Eq. (9), retrieved frequencies $\nu_{fit}$ get independent of $\nu_D$, even for $\nu_D$-values up to 1.3 $\Delta \nu_{Dop}$. Their weighted mean value amounts to 215.823 688 2(6) THz, which is in agreement within $2 \times 10^{-8}$ with the expected value $\nu_0 = 215.823\ 684(3)$ THz, as provided by the HITRAN database [6]. As mentionned, considering the $\tau_D/\Delta t$ values, the frequency step influence is not strictly negligible. Nevertheless the continuous frequency sweeping approximation demonstrates a strong improvement on frequency measurement accuracy even for large $\nu_D$-values. This is simply explain by the fact that the frequency shifts in the real operating mode of the spectrometer (step-by-step sweep) and the continuous frequency sweeping mode approximation remain nearly indistinguishable at the present uncertainty level $((1-a_v).v_D$ is small whatever is $v_D$). For the sake of completeness, numerical simulations have been performed to compare retrieved frequency in the continuous approximation and in the step-by-step numerical analysis. For the present experimental conditions, we conclude that the continuous approximation entails an underestimation of the retrieved frequency of 1.2 MHz (~$6\times10^{-9}$ of the expected frequency), about 2 standard deviations or less than 1/200 of the Doppler width.

To analyse in detail the influence of the frequency constant in step-by-step frequency sweeping mode, complementary experiments have been performed at LPL on ammonia, with fixed values of lock-in time constant (indicated value $\tau_D = 30$ ms) and step duration ($\Delta t = 74.5$ ms) leading to $\tau_D/\Delta t \approx 0.4$. Frequency steps $\Delta v$, in the 0.5-4 MHz range, were positive or negative depending on whether laser frequency was increasing or decreasing, respectively. In such conditions ($0.4<|v_D|<3.3$ MHz), recorded lines are only frequency shifted without any significant line shape distortion. All line fits were done with the usual Voigt profile, the collisional broadening being fixed at the value expected at the sample pressure, the pressure induced line shift being neglected [39]. Corresponding results are resumed in FIG. 4.

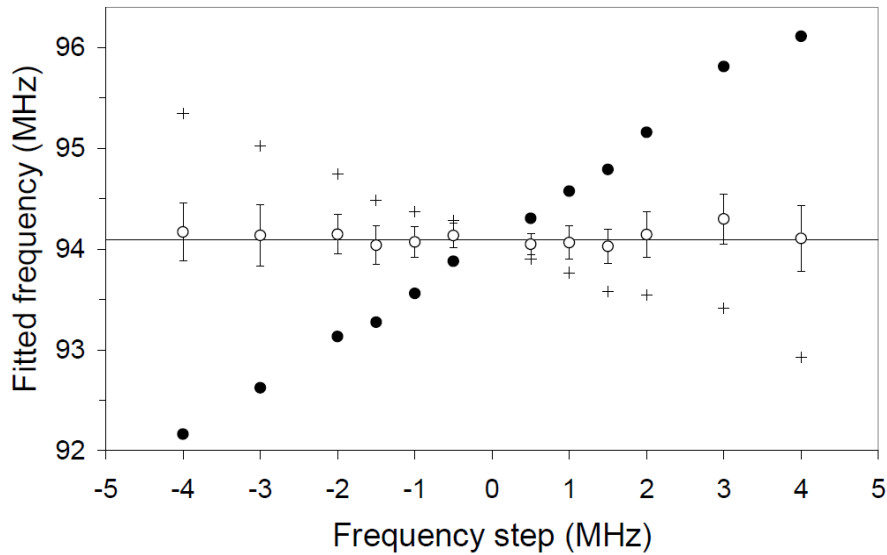

FIG. 4. Influence of the step-by-step mode on the retrieved frequency $v_{fit}$. The frequency of the of the saQ(6,3) rovibrational line of the $v_2$ vibrational mode of $^{14}NH_3$ has been fitted using the Voigt profile (see text): (●) lock-in time constant $\tau_D$ and frequency step $\Delta v$ influences neglected; (+) continuous frequency sweeping approximation; (o) step-by-step frequency sweeping considered with $\tau_D = 36$ ms; (straight line) weighted mean value of retrieved frequencies. Fitted frequency scale is shifted by 28.953 600 THz; step duration $\Delta t = 74.5$ ms; indicated lock-in detection time constant $\tau_D = 30$ ms; temperature =273.15 K; pressure ≈ 1 Torr; cell length = 4.5 cm. For clarity, error bars are 3 standard deviations.

In a first step, line fits were done neglecting influences of lock-in time constant as well as frequency step. Retrieved frequencies $v_{fit}$ (plotted as ●) are actually positively (resp. negatively) shifted according as the laser frequency is increasing (resp. decreasing). If the influence of the time constant $\tau_D$ is considered via Eq. (11), that is assuming a linear evolution of the laser frequency, shifts of retrieved frequencies (plotted as +) are inverted. This feature cannot be explained only by a calibration error in $\tau_D$ (a factor about 2 would be required) and suggests the influence of the step-by-step spectrometer operation. Thus, retrieved frequencies $\nu_{fit}$ have been corrected following numerical simulations (section II.D) and using Eq. (13). It was found that for $\tau_D = 36$ ms, a value 20% larger than the 30 ms indicated one and leading to $a_v = 0.52$, these corrected values (plotted as o) get independent of $\Delta\nu$ steps. The horizontal full line displays the weighted average of these values, 28.953 694 09(8) THz, a result in agreement with the frequency obtained by saturated absorption techniques, 28.953 693 9(1) THz [35]. Moreover, it is worth mentioning that the standard deviations of the set of measurements (0.08 MHz) is comparable to individual $\nu_{fit}$ uncertainty resulting from line fits (between 0.04 and 0.11 MHz). This corresponds to a reproducibility of frequency measurements better than $10^{-8}$ in relative value, which demonstrates the accuracy that can be actually achieved with signal to noise ratios about 200 when the spectrometer apparatus function is properly taken into account.

The artificial frequency shifts discussed in this section are commonly canceled by recording two scans of opposite frequency sweep direction, expected to produce two spectra with opposite shifts about the true resonance value. The latter can therefore be safely recovered. However, this procedure requires a fine control of all experimental parameters during the time needed to record two spectra and a perfect control of the symmetry of the opposite frequency sweeps. The theoretical model described here enables to relax those stringent constraints. Another frequently used trick to get rid of laser frequency drifts consists in randomising the time ordering of the discrete frequencies used to record a spectrum [40] (instead of performing a monotonous step-by-step sweep). However, note that this procedure induces a larger $\nu_D$ mean value, increased by about the number of discret frequencies used. The resulting effect on the measured resonance frequency could certainly be accurately evaluated with an extension of our model, and taken into account in the uncertainty budget.

### B. Collisional broadening measurements

The effects of the detection bandwidth on collisional broadening measurements were analyzed by performing experiments on ammonia using the LPL's spectrometer. Fixing the pressure, a series of scans in both sweep directions were recorded at different frequency constants $\nu_D$ ranging from 0.1 up to nearly 150 MHz. These line shapes were fitted to the Voigt model, with adjustable central frequency, collisional width, line area and base line while fixing the Doppler width at its theoretical value (an usual procedure in collisional broadening measurements). Each dataset was fitted twice: once using a normal Voigt profile, once accounting for the deformations due to the detection bandwidth via Eq. (7) (assuming a continuous laser frequency operation). FIG. 5 shows the measured collisional width as a function of the logarithm of $\nu_D$.

If detection properties are taken in account via Eq. (7) (square symbols), retrieved collisional widths are independent of the frequency constant $\nu_D$, as expected. The standard deviation derived from the fits (about 0.5-1.5%) are also independent of $\nu_D$. If the detection finite bandwidth is ignored (circles), for largest $\nu_D$-values the extracted collisional (*i.e.* Lorentz) widths $\Delta\nu_{Lorentz}$ depart strongly from the expected $\Delta\nu_{col}$ value. Such a divergence, which does not appear explicitly in lowest order terms of Eq. (11), is in fact artificially introduced when fixing the Doppler line width in the fitting procedure. From the experimental results of FIG. 5 which correspond to the Doppler regime ($\Delta\nu_{col} \ll \Delta\nu_{Dop}$), this deviation can be modeled as $\Delta\nu_{Lorentz} = \Delta\nu_{col}\sqrt{1+3(\nu_D/\Delta\nu_{Dop})^2}$. For $\nu_D/\Delta\nu_{Dop} = 1/10$, the corresponding relative broadening is about 1.5%. This means that unless we explicitly account for deformations due to the detection bandwidth, Ref. [22] condition is slightly insufficient in accounting for typical requirements in collisional broadening measurements [41]. Note that increasing and decreasing frequency experiments lead to identical results.

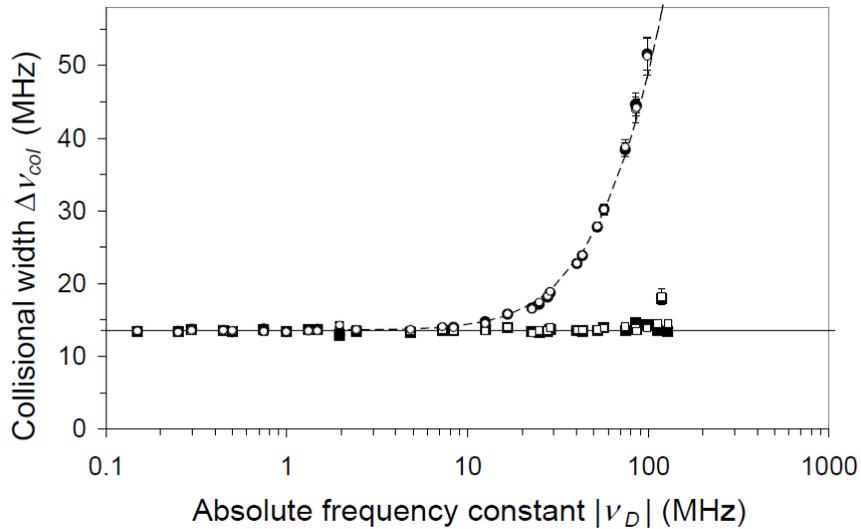

FIG. 5. Collisional broadening of the saQ(6,3) rovibrational line of the $\nu_2$ vibrational mode of $^{14}NH_3$ plotted against the absolute frequency constant (log plot). Spectra were recorded at temperature ≈ 273.15 K and pressure ≈ 1 Torr, for both increasing (open symbols) and decreasing (full symbols) frequencies; fits to a Voigt profile were performed, either by neglecting actual detection properties, circle symbols, or by correcting for them, square symbols (error bars are 3 standard deviations). Full lines correspond to theoretical models (see text for details). Weighted mean value from corrected data is $<\Delta\nu_{col}>$ = 13.59(32) MHz.

The line area behavior was also analyzed using the same experimental data. Within the Ref. [22] condition, $\nu_D/\Delta\nu_{Dop} < 1/10$, the systematic effect induced by the detection bandwidth on line area determination remains smaller than 0.7%.

### C. Line shape analysis

An accurate line shape analysis was performed on water vapor with MPML's spectrometer. FIG. 6 displays a spectrum obtained with the sweeping rate $\dot{\nu}$ = +32.6 MHz/s, and the indicated time constant $\tau_D$ = 3 s, which corresponds to the detection frequency constant $\nu_D$ = 196 MHz, about half the Doppler width. In such conditions, the half-width recording duration $\Delta\nu_{Dop}/\dot{\nu}$ is about 3.7 times larger than the time constant $\tau_D$, to be compared to 20, the recommended minimum value [22]. By comparison with the true line center $\nu_0$, the maximum absorption is shifted by about +180 MHz, nearly half the Doppler width. However, the recorded line looks nearly undistorted at first glance. Several line fits were performed taking account of the Beer-Lambert law. The central absorption frequency $\nu_{fit}$, the collisional line width $\Delta\nu_{col}$, the line area, as well as the parameters of the linear baseline were adjusted whereas $\Delta\nu_{Dop}$ remained fixed at its theoretical value. A fit to the Voigt model but neglecting actual detection properties leads to asymmetric residuals (curve c). Moreover, the retrieved collisional broadening $\Delta\nu_{col} = 93.3(5)$ MHz is about twice larger than the expected value [37].

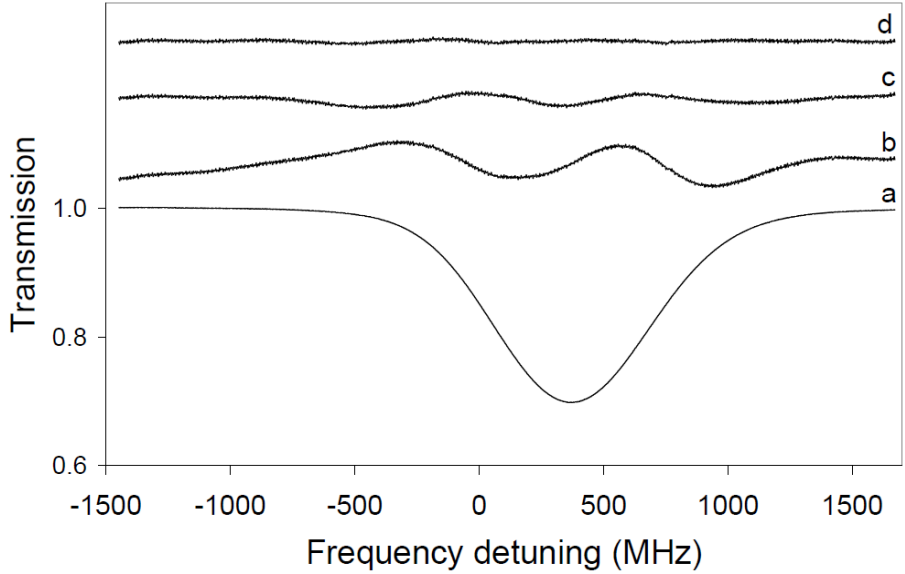

FIG. 6. Transmission of the 7 199.103 cm$^{-1}$ line of H$_2^{18}$O recorded for increasing frequency (see text for details). (a): recorded line. Residuals (magnified by 10): (b): Voigt profile not corrected for detection system properties, $\Delta v_{col}$ = 93.3(5) MHz; (c): corrected Voigt profile, $\Delta v_{col}$ = 41.7(2) MHz; (d): corrected SD-Galatry profile, $\Delta v_{col}$ = 49.78(12) MHz, achieved signal to noise ratio ≈ 4 500. Frequency scale (in MHz) is shifted by 215.823 500 THz; Temperature = 296.073 0(4) K; pressure ≈ 3.6 Torr; cell length ≈ 30 cm; frequency step = 3 MHz; step duration = 92 ms; indicated lock-in time constant $\tau_D$ = 3 s.

The ratio $\tau_D/\Delta t$ has been chosen to be large enough (equal to 32) so that the continuous frequency sweeping mode approximation is valid. It has been applied by fitting line shapes using Eq. (9). Accounting for detection properties leads to a much better result (curve d), but still having the well known W structure in the residuals, which is characteristic of a line narrowing effect [12]. According to [37], this departure from the Voigt profile results mainly from the dependence of collisional broadening on molecular speeds with a small contribution of the Dicke effect via velocity-changing collisions. So, a nearly perfect result (curve e) has been obtained using the symmetric version of the speed-dependent Galatry profile (SDGP). In this fit, the diffusion parameter was fixed ($\beta_{Gal}$ = 0.4 MHz/Torr [37]) whereas speed dependent effects were taken into account via the hypergeometric model [37] in which the collisional broadening and corresponding velocity exponent were adjusted, leading to $\Delta v_{col} = 49.73(12)$ MHz and $m = 0.64(1)$. These values, even being obtained from a single recording performed with quite unusual detection properties are reasonably good. In particular, the $m$-value is close to the 0.5 value given in Ref. [34].

## V. DOPPLER BROADENING THERMOMETRY

One of the motivations of the present work is to further gain in the capability to make precision spectroscopic measurements devoted to metrological applications. In this section a particular emphasis is put on an accurate spectroscopic determination of the Boltzmann constant by DBT. In proof-of-principle experiments performed on NH$_3$ and CO$_2$ molecules, $k_B$ was determined with a combined uncertainty of 190 and 160 parts per million (ppm), respectively [42,43]. At the time, the spectral analysis was performed using simplified models such as gaussian and Voigt profiles. In the last few years, an ambitious goal was set to reach a target uncertainty of 1 ppm, needed for the new definition of the kelvin unit [17-19]. The required technical improvements and upgrades of the spectrometers have been accompanied by an increasingly refined interpolation of the experimental profiles [44-47]. Particularly, a great attention is being paid to the role of speed-dependent effects [39, 48]. Also, other molecules and atoms such as Rb, $H_2^{18}O$ or $C_2H_2$ are now probed [32, 49, 50]. Finally, there are significant efforts to produce a complete uncertainty budget, which includes the understanding and possible reduction of sources of systematical deviations [33, 34].

Since the target uncertainty is in the 1 ppm range, a reliable quantification of the finite bandwidth contribution to the uncertainty budget is necessary. This is obtained through both experiments and numerical simulations based on the theoretical model of section II.

### A. Doppler width measurements

In order to confirm theoretical predictions of section II, the set of measurements recorded with MPML's spectrometer and analysed in section IV.A has been once more fitted considering the continuous frequency sweeping model (a suitable approximation for this particular set of data, see section IV.A) and using the speed dependent Galatry profile. Unlike previous analyses, the Doppler width is now considered as a free parameter and collisional broadening $\Delta v_{col}$, pressure-induced shift, velocity exponent $m$ and the diffusion parameter $\beta_{Gal}$, as fixed parameters. Note first that we draw the same broad conclusions as those outlined in section IV.A for the central absorption frequency $v_{fit}$. Concerning Doppler width measurements, the results are summarised in FIG. 7 where the retrieved gaussian widths $\Delta v_{Gauss}$ are plotted against the logarithm of the frequency constants $v_D$. When neglecting the influence of detection properties, the gaussian widths, $\Delta v_{Gauss}$, look weakly affected by the frequency constant up to 30 MHz, a tenth of the Doppler width. The retrieved gaussian widths can be compared to the results predicted by Eq. 12. This approximate theory, valid for small $v_D$-values, is in agreement with observations up to $v_D$ of about 100 MHz, *i.e.* 1/3 the Doppler width (see the dotted curve of FIG. 7). Above this value, the extracted values deviate due to line asymmetries.

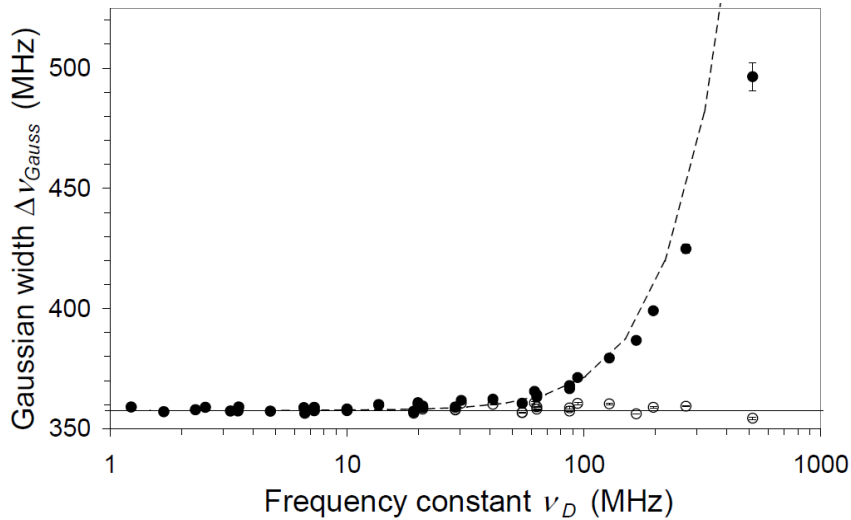

FIG. 7. Doppler width of the 7,199.103 cm$^{-1}$ line of $H_2^{18}O$ retrieved from a SD-Galatry profile. Fitted gaussian widths $\Delta v_{Gauss}$ are plotted versus the frequency constant $v_D$ (log plot): (●) uncorrected values; (o) corrected values for the detection system properties (see text for details). Approximate theoretical uncorrected values: dotted line. Weighted mean value of corrected data (full line): $<\Delta v_{Gauss}>$ = 359.1(15) MHz. Error bars are 3 standard deviations.

By contrast, if the detection properties are taken into account in the line fitting procedure, the retrieved Doppler widths are clearly independent on $v_D$, even for the largest values of the frequency constant. In this case, a weighted mean leads to 359.1(15) MHz, in agreement to within 6 10$^{-3}$ with the expected Doppler width, a good result for such unusual detection conditions for time constant and/or frequency sweeping rate. Once again the continuous frequency sweeping approximation demonstrates a strong improvements of line width measurement accuracy.

Note that experiments at LPL (section IV.B) also lead to similar conclusions, namely that the retrieved gaussian width is in agreement to within less than 1% with the expected Doppler width when detection properties are considered via Eq. (9).

### B. Application to the Boltzmann constant determination

As mentioned before, the primary motivation for this paper was to model the detection systems in DBT experiments for measuring the Boltzmann constant. So far, we have shown that the finite bandwidth of a detection system can cause distortions to the measured lineshapes, and we have introduced a model to account for these effects. Therefore, one possible approach for the spectroscopic determination of $k_B$ from here onwards would be to include this model in the spectral analysis. This would make the fitting procedure even more complicated than it is presently [33, 34]. Furthermore, very precise measurements of the filter characteristics would be required, without which the final uncertainty would increase significantly. We therefore propose a simpler alternative: to operate in a region of the parameters' space where the model indicates that the effects are negligible. The purpose of this section is to determine this favorable region of the parameters' space. We note in advance that there will be competing interests: the level of signal deformation and the signal-to-noise ratio that both increase with $\nu_D$.

We conduct numerical simulations, similar to those outlined in section II.D for the determination of the line center except this time, we concern ourselves with the width. We generate fake distorted gaussian lineshapes, which we fit using a pure gaussian function. Note that the use of a gaussian profile is valid, as collisional broadening is minimal under usual operating conditions for $k_B$ determination. We then plot the deviation of the measured width from its true value $\left(\Delta\nu_{Gauss} - \Delta\nu_{Dop}\right)$ as a function of $\nu_D$, the frequency constant. This is shown on FIG. 8 for the 2$^{nd}$ order filter cases $Q$: 1/2 and $\sqrt{1/2}$, the roll-off being -12dB/oct. Once again, all values have been normalized to the nominal Doppler width. Note the target error tolerance, highlighted by a horizontal line at the 1 ppm level so our final $k_B$ measurements must be made in conditions where the deviation is below this line. The results of numerical simulations are shown together with the analytical result assuming a continuous frequency sweep.

First consider the $Q = 1/2$ case. For the continuous frequency sweeping model, the relative deviations of the retrieved Doppler width, given by $\left(\Delta\nu_{Gauss} - \Delta\nu_{Dop}\right)/\Delta\nu_{Dop} \approx 0.5\left(\nu_D/\Delta\nu_{Dop}\right)^2$ as derived from Eq. (12), is drawn as a thin dotted line. We see that the target accuracy of 1 ppm will be reached if $\nu_D/\Delta\nu_{Dop}$ is smaller than $\approx 1.5 \times 10^{-3}$. The associated numerical simulations show a more interesting result, which depends strongly on the frequency step $\Delta\nu$ and on the ratio $\tau_D/\Delta t$. Therefore, curves corresponding to two $\Delta\nu$-values have been drawn. As expected, in case of large $\tau_D/\Delta t$ ratios, the model is well approximated by the continuous frequency sweep. However, if this ratio gets smaller than about 0.35, the deviation from the Doppler width is strongly reduced. Although this may seem promising, this happens because the time constant $\tau_D$ is short, implying a reduction in the signal-to-noise ratio. Nevertheless, ignoring this last difficulty and from examination of FIG. 8, we conclude that for a $Q = 1/2$ filter, deviations are reduced to an acceptable level for example with a frequency constant $\nu_D$ about 700 smaller than $\Delta\nu_{Dop}$, frequency steps $\Delta\nu$ as small as 1/330 of the Doppler width $\Delta\nu_{Dop}$, which implies a ratio $\tau_D/\Delta t$ smaller than 0.25. Note that in the 1$^{st}$ order filter case (-6dB/oct roll-off), numerical simulations lead to the same kind of requirements, if $\nu_D$ is chosen about 1000 smaller than $\Delta\nu_{Dop}$.

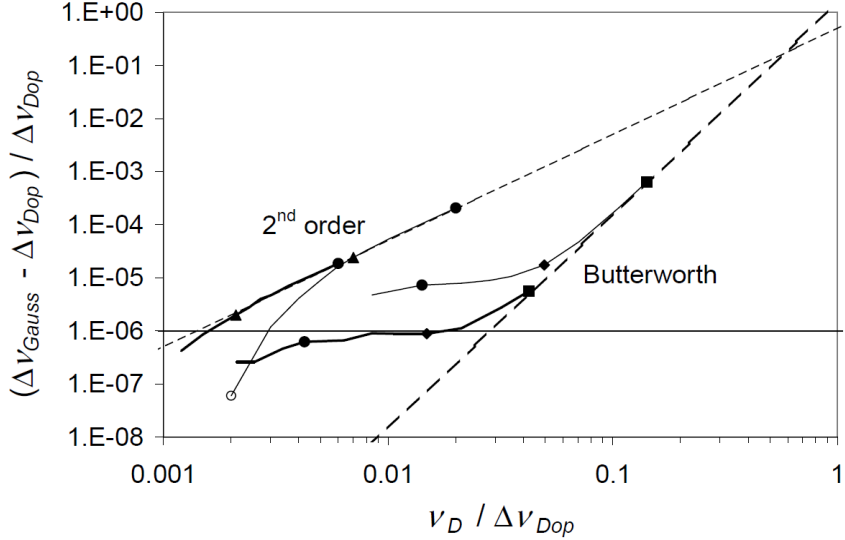

FIG. 8. Influence of frequency sweeping conditions on the Doppler broadening determination (2$^{nd}$ order filters with $Q = 1/2$ and $\sqrt{1/2}$ (Butterworth case)). Relative broadening deviations are plotted versus frequency constant $\nu_D/\Delta\nu_{Dop}$. Frequency step $\Delta\nu/\Delta\nu_{Dop} = 0.003$ (thick line) and 0.01 (thin line) are drawn. For convenience, some $\tau_D/\Delta t$ ratios are specified by symbols: (○) 0.1; (▲) 0.35; (●) 1.0; (♦) 3.5; (■) 10. Thin and thick dotted lines refer to continuous frequency sweeping models ($Q = 1/2$ and $\sqrt{1/2}$, respectively). The 1 ppm target is depicted by the horizontal solid line.

The analytical model for the Butterworth 2$^{nd}$ order filter with a −12 dB/oct roll-off and $Q = \sqrt{1/2}$ is also shown on FIG. 8 as a bold dashed line. This case is clearly more attractive because Eq. (12) predicts a gaussian width which is independent of $\nu_D$ to lowest order. As a matter of fact, for the continuous frequency sweeping case, numerical simulations show this width can be modeled as $(\Delta\nu_{Gauss} - \Delta\nu_{Dop})/\Delta\nu_{Dop} \approx 1.5\,(\nu_D/\Delta\nu_{Dop})^4$, a 4$^{th}$ order polynomial dependence instead of the 2$^{nd}$ order dependence of the previous cases. By contrast with the $Q = 1/2$ 2$^{nd}$ order filter, the continuous frequency sweeping model appears to be a lower limit: decreasing the $\tau_D/\Delta t$ ratios no longer reduces the observed deviations. A stronger reduction is only obtained by reducing the step size. Nonetheless, the 1 ppm target is reached with a $\Delta\nu$ about 330 times smaller than $\Delta\nu_{Dop}$ but the $\tau_D/\Delta t$ ratio can be as large as 3.5, more than one order of magnitude larger than in the case of the $Q = 1/2$ 2$^{nd}$ order filter. This demonstrates the advantage of a Butterworth filter as it would allow the use of a much larger time constants $\tau_D$ or sweep speeds. This implies that for a fixed acquisition time, the experiment would yield better signal-to-noise ratio, so reduced experimental uncertainty, using a Butterworth filter.

It is worth noting that the recent measurement of $k_B$ at the MPML [49] was carried out under very favourable conditions, using a first order low pass filter so that $\nu_D$ was about 2 kHz (namely, five orders of magnitude smaller than the Doppler width), while the ratio $\tau_D/\Delta t$ was set to $\approx 1.6 \times 10^{-3}$. These conditions make it possible to neglect the detection bandwidth effect in the uncertainty budget [49]. On the other hand, this contribution is larger than 10 ppm in the determination of $k_B$ reported in refs [46, 35], $v_D/\Delta v_{Dop}$ being about $7 \times 10^{-3}$ and the ratio $\tau_D/\Delta t$ being set to $\approx 0.35$. This source of inaccuracy not considered at that time remains below the reported combined standard uncertainty of 50 ppm [13]. Nevertheless, for this project, the acquisition conditions will have to be reconsidered in order to reduce the impact of the limited detection bandwidth to below 1 ppm for future measurements.

Note that the method mentioned in section IV.A, consisting in randomising the time ordering of frequencies to get rid of frequency drifts is particularly detrimental when measuring a line-width and thus $k_B$, as it induces a much larger $v_D$ mean value and in turn a larger impact on the recorded width, and especially the Doppler width.

## VI. CONCLUSION

Precision measurements in laser spectroscopy require high signal-to-noise ratio, on the one hand, and high spectral fidelity, on the other one. A reduced detection bandwidth may have significant consequences on the observed line shape, adding asymmetries, shifting its central frequency or broadening the line. A theoretical model has been proposed for a careful analysis of these effects. In case of a continuous evolution of the laser frequency, the line shape can be set in a quasi-analytical form that allows one to consider the influence of the detection bandwidth, thus avoiding possible systematical deviations in the retrieved parameters. Numerical simulations performed in the case of a step-by-step frequency scan lead to an empirical expression for the correction to be applied to the center frequency retrieved from a fit of a distorted profile.

The model has been severely and accurately validated by applying it to the analysis of high-quality molecular spectra. The impact of the detection bandwidth could be quantified for various line shape parameters: Doppler broadening, collisional broadening, the dependence of the latter on molecular speeds, Dicke narrowing contribution.

The dedection bandwidth induced line shape distorsion and the resulting inaccuracy in the measured parameters may impact many fields of research, from atmospheric and interstellar physics to precision spectroscopic measurements devoted to metrological applications, tests of quantum electrodynamics or other fundamental laws of nature. As an example, emphasis has been put on the repercussions on the precise determination of the Boltzmann constant by the Doppler Broadening Thermometry. Our study allows us to work out the optimum experimental conditions (integration time, frequency scan rate, type of filter, and frequency step) required to reach the targeted uncertainty of 1 ppm level of accuracy. The present study could easily be extended to other precision spectroscopic measurements in order to quantify and possibly reduce the resulting inaccuracy potentially affecting such experiments.

## AKNOWLEDGMENTS

The authors acknowledge financial support from CNRS, Université Paris 13 and LNE. This work is part of projects NCPCHEM (2010 BLAN 724 3) and CaPPA (ANR-10-LABX-005 through the Programme d'Investissement d'Avenir), both funded by the Agence Nationale de la Recherche (ANR, France). François Rohart thanks Bernard Ségard for helpful discussions.

22[50] K. M. T. Yamada, A. Onae, F. -L. Hong, H. Inaba, H. Matsumoto, Y. Nakajima, F. Ito, and T. Shimizu, J. Mol. Spectrosc. **249**, 95 (2008)